\documentclass[aip,pop,10pt]{revtex4-1}
\pdfoutput=1

\usepackage{amsmath}
\usepackage{amssymb}
\usepackage{graphicx}
\usepackage{textcomp}

\begin{document}
\title{Wave dispersion in the hybrid-Vlasov model: verification of Vlasiator}

\author{Yann Kempf}
\email[]{yann.kempf@fmi.fi}
\affiliation{Finnish Meteorological Institute, P.O.\ Box 503, 00101 Helsinki,
Finland}
\affiliation{University of Helsinki, Department of Physics, P.O.\ Box 64,
00014 Helsinki, Finland}

\author{Dimitry Pokhotelov}
\affiliation{Finnish Meteorological Institute, P.O.\ Box 503, 00101 Helsinki,
Finland}
\affiliation{University of Helsinki, Department of Physics, P.O.\ Box 64,
00014 Helsinki, Finland}

\author{Sebastian von Alfthan}
\affiliation{Finnish Meteorological Institute, P.O.\ Box 503, 00101 Helsinki,
Finland}

\author{Andris Vaivads}
\affiliation{Swedish Institute of Space Physics, Box 537, 751 21 Uppsala,
Sweden}

\author{Minna Palmroth}
\affiliation{Finnish Meteorological Institute, P.O.\ Box 503, 00101 Helsinki,
Finland}

\author{Hannu E.\ J.\ Koskinen}
\affiliation{Finnish Meteorological Institute, P.O.\ Box 503, 00101 Helsinki,
Finland}
\affiliation{University of Helsinki, Department of Physics, P.O.\ Box 64,
00014 Helsinki, Finland}

\date{\today}

\begin{abstract}
Vlasiator is a new hybrid-Vlasov plasma simulation code aimed at simulating the
entire magnetosphere of the Earth. The code treats ions (protons) kinetically
through Vlasov's equation in the six-dimensional phase space while electrons are
a massless charge-neutralizing fluid
[M. Palmroth \textit{et al.}, Journal of Atmospheric and Solar-Terrestrial
Physics \textbf{99}, 41 (2013); A. Sandroos \textit{et al.}, Parallel Computing
\textbf{39}, 306 (2013)].
For first global simulations of the magnetosphere, it is critical to verify and
validate the model by established methods. Here, as part of the verification of
Vlasiator, we characterize the low-$\beta$ plasma wave modes described by this
model and compare with the solution computed by the Waves in Homogeneous,
Anisotropic Multicomponent Plasmas (WHAMP) code
[K. R\"onnmark, Kiruna Geophysical Institute Reports \textbf{179} (1982)],
using dispersion curves and surfaces produced with both programs. The match
between the two fundamentally different approaches is excellent in the
low-frequency, long wavelength range which is of interest in global
magnetospheric simulations. The left-hand and right-hand polarized wave modes
as well as the Bernstein modes in the Vlasiator simulations agree well with the
WHAMP solutions. Vlasiator allows a direct investigation of the importance of
the Hall term by including it in or excluding it from Ohm's law in simulations.
This is illustrated showing examples of waves obtained using the ideal Ohm's law
and Ohm's law including the Hall term. Our analysis emphasizes the role of the
Hall term in Ohm's law in obtaining wave modes departing from ideal
magnetohydrodynamics in the hybrid-Vlasov model.
\vspace{10pt}

Copyright 2013 American Institute of Physics. This article may be downloaded
for personal use only. Any other use requires prior permission of the author and
the American Institute of Physics.

The following article appeared in Physics of Plasmas \textbf{20}, 112114 (2013)
and may be found at \url{http://dx.doi.org/10.1063/1.4835315}.
\end{abstract}

\maketitle
\section{Introduction}
The exponential growth in available computing power has made hybrid and fully
kinetic plasma simulations increasingly feasible for a variety of space plasma
applications. Non-exhaustive examples include hybrid particle-in-cell
(hybrid-PIC) simulations of planetary environments
\cite{Sillanpaa2007,BrechtJGR1991} and magnetospheres
\cite{LinJGR2003,LinWangJGR2005,Omidi2005,omelchenko2012}, full-PIC studies of
magnetic reconnection \cite{Divin2012,Pritchett2013}, local hybrid-Vlasov
simulations of wave-particle interactions in the solar wind \cite{Valentini2010}
or full-Vlasov simulations of the Kelvin-Helmholtz instability \cite{Umeda2010}.
Vlasiator is a new self-consistent hybrid-Vlasov simulation code in which the
ions (protons) are treated kinetically via Vlasov's equation and electrons are a
massless charge-neutralizing fluid. It is based on a robust finite volume
method, which has been optimized for the modeling of the entire magnetosphere of
the Earth\cite{Sandroos2013,Palmroth2012} (\url{http://vlasiator.fmi.fi}). To
our knowledge, Vlasiator is the first hybrid-Vlasov code to allow simulations on
this scale. An important advantage of the hybrid-Vlasov model with respect to
(hybrid-)PIC methods is the absence of noise related to the low number of
particles representing the distribution function, as hybrid-Vlasov algorithms
propagate the full distribution function in the six-dimensional phase space
using Vlasov's equation. The uniform sampling in velocity space provides a
description of the distribution function with a quality comparable to spacecraft
measurements. However, the six-dimensional representation of the distribution
function in the hybrid-Vlasov approach makes the memory and computing
requirements high, even for modern massively parallel supercomputers.

The hybrid-Vlasov scheme is a relatively new approach to computational plasma
physics at large scales because of the aforementioned need of computing
resources. Hence it is critical to provide basic benchmarks to assess the
possibilities and limits of the model in terms of its physical features. As part
of the verification of Vlasiator we perform local simulations to investigate the
propagation of low-$\beta$ plasma waves in the hybrid-Vlasov model by studying
wave dispersion in a variety of cases. The dispersion of the left- and
right-hand polarized modes as well as the ion-acoustic waves propagating
parallel to the magnetic field has been used before to verify a hybrid-Vlasov
simulation code\cite{Valentini2007}. We extend this test to include more plasma
wave modes in all propagation directions ranging from parallel to perpendicular
to the magnetic field and present dispersion surface plots. The cases included
here are in a range of parameters relevant to understand global magnetospheric
simulations. Thus this study helps bolstering the interpretation of current and
future large-scale simulations of the magnetosphere of the Earth in 2+3 and 3+3
spatial and velocity dimensions. We compare the results obtained with Vlasiator
to the solutions calculated using the Waves in Homogeneous, Anisotropic
Multicomponent Plasmas code (WHAMP, \url{https://github.com/irfu/whamp}), which
solves the linearized kinetic dispersion equation
numerically\cite{Ronnmark1982,Ronnmark1983}. The simple and robust solution
approach of WHAMP starting from the general dispersion equation and its strong
establishment as a tool for the determination of wave dispersion in homogeneous
plasmas make it an ideal tool to verify Vlasiator results.

Since the central interest of Vlasiator lies in simulating the entire
magnetosphere of the Earth, we focus on wave modes at spatial and temporal ion
scales. In ideal magnetohydrodynamic (MHD) theory the only possible wave modes
are the Alfv\'en wave, which is a shear electromagnetic mode propagating at all
angles except perpendicularly to the magnetic field, and the fast and slow
magnetosonic modes, which are compressional electromagnetic modes propagating at
an oblique angle with respect to the magnetic field\cite{KoskinenSpaceStorms}.
The dispersion equation of the shear Alfv\'en wave is given by
\begin{equation}
   \frac{\omega}{k} = v_\mathrm{A}\cos\theta,
\label{eq:Disp_MHD_Alfven}
\end{equation}
where $\omega$ is the angular frequency, $k$ the wave number, $v_\mathrm{A} =
B/\sqrt{\mu_0\rho_m}$ ($B$: magnetic field intensity, $\mu_0$: permeability of
vacuum, $\rho_m$: mass density) is the Alfv\'en speed and $\theta$ is the angle
between the wave vector $\mathbf{k}$ and the magnetic field vector $\mathbf{B}$.
The fast ($+$) and slow ($-$) magnetosonic wave dispersion equation is given by
\begin{equation}
   \left(\frac{\omega}{k}\right)^2 =
   \frac{1}{2}\left(v_\mathrm{s}^2 + v_\mathrm{A}^2\right) \pm \frac{1}{2}\sqrt{
   \left(v_\mathrm{s}^2 + v_\mathrm{A}^2\right)^2 -
   4v_\mathrm{s}^2v_\mathrm{A}^2\cos^2\theta},
\label{eq:Disp_MHD_MS}
\end{equation}
where $v_\mathrm{s} = \sqrt{k_\mathrm{B} T / m_\mathrm{i}}$ ($k_\mathrm{B}$:
Boltzmann constant, $T$: temperature, $m_\mathrm{i}$: ion mass) is the sound
speed. In a low-$\beta$ plasma we have $v_\mathrm{A} > v_\mathrm{s}$. For
parallel propagation ($\theta = 0$), the Alfv\'en and fast magnetosonic modes
coincide and the mode propagates at $v_\mathrm{A}$, whereas the slow
magnetosonic mode becomes the sound wave propagating at $v_\mathrm{S}$. In the
perpendicular case ($\theta = \pi/2$), the Alfv\'en and slow mode cannot
propagate and only the fast magnetosonic mode propagating at magnetosonic speed
$v_\textrm{MS}=\sqrt{v_\mathrm{A}^2 + v_\mathrm{s}^2}$ subsists. These modes are
dispersionless.

Multi-fluid and kinetic theories describe more plasma wave modes. The left- and
right-hand polarized modes (L- and R-mode hereafter) propagate along or
quasi-parallel to the magnetic field. They couple to the ion and electron
gyromotion, therefore they resonate at the ion and electron gyrofrequency,
respectively. In the low frequency, small wave number limit both modes converge
towards the ideal MHD Alfv\'en mode.

The last wave modes in the range of frequencies and wavelengths of interest in
this paper are the ion Bernstein modes, which are electrostatic ion-cyclotron
resonances (quasi-)perpendicular to the magnetic field. Their dispersion
equation can be expressed in terms of infinite sums of Bessel functions in the
hybrid-Vlasov model\cite{Kazeminezhad1992}. A thorough review of plasma wave
dispersion surfaces obtained with WHAMP at ion and electron scales (including
the modes introduced in this section) is given by \citet{Andre1985}.

This paper is organized as follows. In Section \ref{sec:Models_methods} we
present Vlasiator and WHAMP as well as the simulation setup used with Vlasiator.
Then we present our results in Section \ref{sec:Results}. We show the importance
of the Hall term in Ohm's law, and we study dispersion curves for parallel,
perpendicular and oblique propagation as well as dispersion surfaces. We discuss
the results in Section \ref{sec:Discussion} and conclusions are drawn in Section
\ref{sec:Conclusions}.

\section{Model and methods}
\label{sec:Models_methods}
\subsection{The hybrid-Vlasov model in Vlasiator}
The solver for Vlasov's equation implemented in the hybrid-Vlasov model
of Vlasiator is based on a three-dimensional wave propagation
algorithm\cite{Leveque1997,Langseth2000} applied separately to translation in
position space and acceleration in velocity space. It is self-consistently
coupled to a field solver\cite{LondrilloDelZanna2004} which uses divergence-free
magnetic field reconstruction\cite{Balsara2009}. We only introduce here the
solver features relevant to this study.

Vlasiator solves Vlasov's equation
\begin{equation}
\frac{\partial}{\partial t}f(\mathbf{r},\mathbf{v},t)+
\mathbf{v}\cdot\nabla_{r}f(\mathbf{r},\mathbf{v},t)+
\mathbf{a}\cdot\nabla_{v}f(\mathbf{r},\mathbf{v},t)=0,
\label{eq:vlasov}
\end{equation}
where $\mathbf{r}$ and $\mathbf{v}$ are the spatial and velocity coordinates,
$t$ is time, $f(\mathbf{r},\mathbf{v},t)$ is the six-dimensional phase-space
density of ions with mass $m$ and charge $q$, and acceleration $\mathbf{a}$
is due to the Lorentz force
\begin{equation}
\mathbf{a}=\frac{q}{m}(\mathbf{E}+\mathbf{v}\times\mathbf{B}),
\label{eq:Lorentz}
\end{equation}
in which $\mathbf{E}$ is the electric field and $\mathbf{B}$ is the magnetic
field.

In the hybrid-Vlasov model Vlasov's equation is coupled to Maxwell's equations.
The displacement current is neglected in the Amp\`{e}re-Maxwell law and the
equations take the form
\begin{align}
\nabla\times\mathbf{E} & =
-\frac{\partial}{\partial{t}}\mathbf{B},\label{eq:maxwell-faraday}\\
\nabla\times\mathbf{B} & =
\mu_{0}\mathbf{j},\label{eq:maxwell-ampere}
\end{align}
where $\mathbf{j}$ is the total current density. Note that the equation
$\nabla\cdot\mathbf{B} = 0$ is respected by the field propagation algorithm of
Vlasiator by construction, provided the initial conditions are
divergence-free\cite{LondrilloDelZanna2004,Balsara2009}.

Ohm's law describes the relationship between the electric and the magnetic
field. It is needed to close the hybrid-Vlasov system of equations, when
updating the magnetic field using Faraday's law (Eq.\ \ref{eq:maxwell-faraday}).
In the present study Vlasiator is using the ideal Ohm's law supplemented by a
Hall term (rightmost term in Eq.\ \ref{eq:ohms-law}) with first-order spatial
accuracy:
\begin{equation}
\mathbf{E}=-\mathbf{V}_{\mathrm{i}}\times\mathbf{B} +
\frac{1}{\rho_{q}}\mathbf{j}\times\mathbf{B}.
\label{eq:ohms-law}
\end{equation}
The ion charge density $\rho_{q}$ and (ion) bulk velocity
$\mathbf{V}_\mathrm{i}$ are obtained from velocity moments of the distribution
function, $\mathbf{j}$ is computed using Eq.\ \eqref{eq:maxwell-ampere}. This
effectively represents electrons as a massless, charge-neutralizing fluid. The
rest of the solvers in Vlasiator retain second-order spatial accuracy throughout
in smooth cases. In cases with strong spatial gradients---in position or
velocity space alike---flux limiters effectively reduce the spatial accuracy of
the solvers in order to preserve the numerical stability of the scheme.

Note that although the Hall term in the computation of the electric field can be
neglected using the ideal Ohm's law, it has to be retained in the electric field
input into the Lorentz force (Eq.\ \ref{eq:Lorentz}) in order to model bulk
forces on the ions. This makes Vlasiator a proper Hall-less hybrid model when
the Hall term in Ohm's law is not used\cite{karimabadi2004}.

\subsection{The WHAMP code}
WHAMP is a code solving the linear analytic dispersion equation of waves in
magnetized plasmas. It can include several populations with
differing number density, mass, temperature, loss cone, anisotropy and drift
parameters for anisotropic Maxwellian
distributions\cite{Ronnmark1982,Ronnmark1983,Andre1985}. A general form of the
plasma wave dispersion equation is
${\mathbf{D}\left(\omega,\mathbf{k}\right)\cdot\mathbf{E}\left(\omega,
\mathbf{k}\right) = 0}$,
where $\mathbf{D}$ is the dispersion tensor and $\mathbf{E}$ the wave electric
field. Solutions can be found by equating the determinant of the dispersion
tensor to zero, $\left|\mathbf{D}\left(\omega,\mathbf{k}\right)\right| = 0$.
WHAMP solves this in a linearized form using a Pad\'e approximant to ensure a
fast computation yet good approximation of the result.

The WHAMP interface is designed to take in the plasma parameters initially and
then allow to query the solution point by point in the ($\mathbf{k}$-$\omega$)
space. Given an initial ($k_\perp$,$k_\parallel$) point WHAMP tries to find a
wave mode close-by and returns the frequency, wave vector and growth rate of the
mode among others. A script querying WHAMP systematically is used to ease
comparison with the full dispersion plots and surfaces obtained with Vlasiator
following the methods presented in the next section.

When comparing WHAMP and Vlasiator results, the electron temperature
$T_\mathrm{e}$ in WHAMP is set to a small value in order to mimic the absence of
electron pressure gradient effects in our hybrid-Vlasov model. The electron
temperature is ignored in Vlasiator, effectively suppressing in the model the
ion-acoustic wave for which $T_\mathrm{e}\gg T_\mathrm{i}$ must hold.

\subsection{Simulation setup and processing}
The simulation setup of Vlasiator in the present study consists of a
one-dimensional spatial domain along the $x$-axis, constraining the wave vector
to be in that direction, with fully periodic boundary conditions. The angle
between the magnetic field and the wave vector is defined by setting the
magnetic field orientation with respect to the simulation box. The initial
conditions are uniform up to small random perturbations in the number density
and bulk velocity. The velocity distribution is isotropic and Maxwellian,
therefore excluding waves growing from anisotropy-driven instabilities. The
system relaxes and no forcing is applied during the simulation.

A two-dimensional space-time ($x$-$t$) dataset is formed by saving the spatial
profile of a bulk variable at every (constant) time step. The total run time is
typically several ion gyroperiods. The data is first windowed along the time
dimension using a Hamming window to reduce the noise induced by the abrupt
start and stop of the time series. Then it is subjected to a discrete Fourier
transformation to produce a ($k$-$\omega$) dispersion plot.

Dispersion surfaces are computed by extracting the points above a threshold in
each ($k$-$\omega$) dataset to retrieve the dispersion branches for multiple
angles. The extracted data is interpolated and re-sampled on a Cartesian
coordinate grid to obtain smoother dispersion surfaces more readily comparable
with the dispersion surfaces from WHAMP.

The simulation and plasma parameters of all Vlasiator simulations presented in
this paper are given in Table \ref{tab:Simulation_parameters}. WHAMP uses
exactly the same parameters as input, except for the electron temperature as
discussed above. 

\begin{table} 
   \centering
   \caption{Parameters of the Vlasiator simulations presented in this paper. The
   propagation angles with respect to the magnetic field are $\theta=0.001$ for
   the parallel cases without and with Hall term (Fig.~\ref{fig:Par});
   $\theta=1.57$ for the perpendicular case (Fig.~\ref{fig:Perp});
   $\theta=0.001,0.1,0.3,0.5,0.7,1.0,1.2,1.3,1.4,1.57$ for the dispersion
   surfaces (Fig.~\ref{fig:Surfaces}); and $\theta=0.3$ for the oblique case
   (Fig.~\ref{fig:Oblique}).}
   \begin{tabular}{lcc}
      \hline\hline
      Simulation parameters & Dispersion surface/parallel/oblique &
      Perpendicular \\
      Simulation domain length & $2.5\cdot10^8$~m & $5.0\cdot10^8$~m \\
      Number of spatial cells & $10{,}000$ & $10{,}000$ \\
      Velocity space resolution &$4.0\cdot10^3$~m/s & $4.0\cdot10^3$~m/s \\
      Time step & $0.001$~s & $0.025$~s \\
      \hline
      Plasma parameters & & \\
      Ion (proton) number density &
      $1.0\cdot10^6$~m$^{-3}$ & $1.0\cdot10^4$~m$^{-3}$ \\
      Ion (proton) temperature & $1.0\cdot10^5$~K & $1.0\cdot10^5$~K \\
      Magnetic field intensity & $5.0\cdot10^{-8}$~T & $1.0\cdot10^{-9}$~T \\
      Plasma $\beta$& $1.4\cdot10^{-3}$ & $3.5\cdot10^{-2}$ \\
      \hline\hline
   \end{tabular}
\label{tab:Simulation_parameters}
\label{tab:Plasma_parameters}
\end{table}

\section{Results}
\label{sec:Results}

\subsection{The Hall term in Ohm's law and parallel propagation}
\label{sec:Parallel}
\begin{figure}
   \centering
   \includegraphics[width=\textwidth]{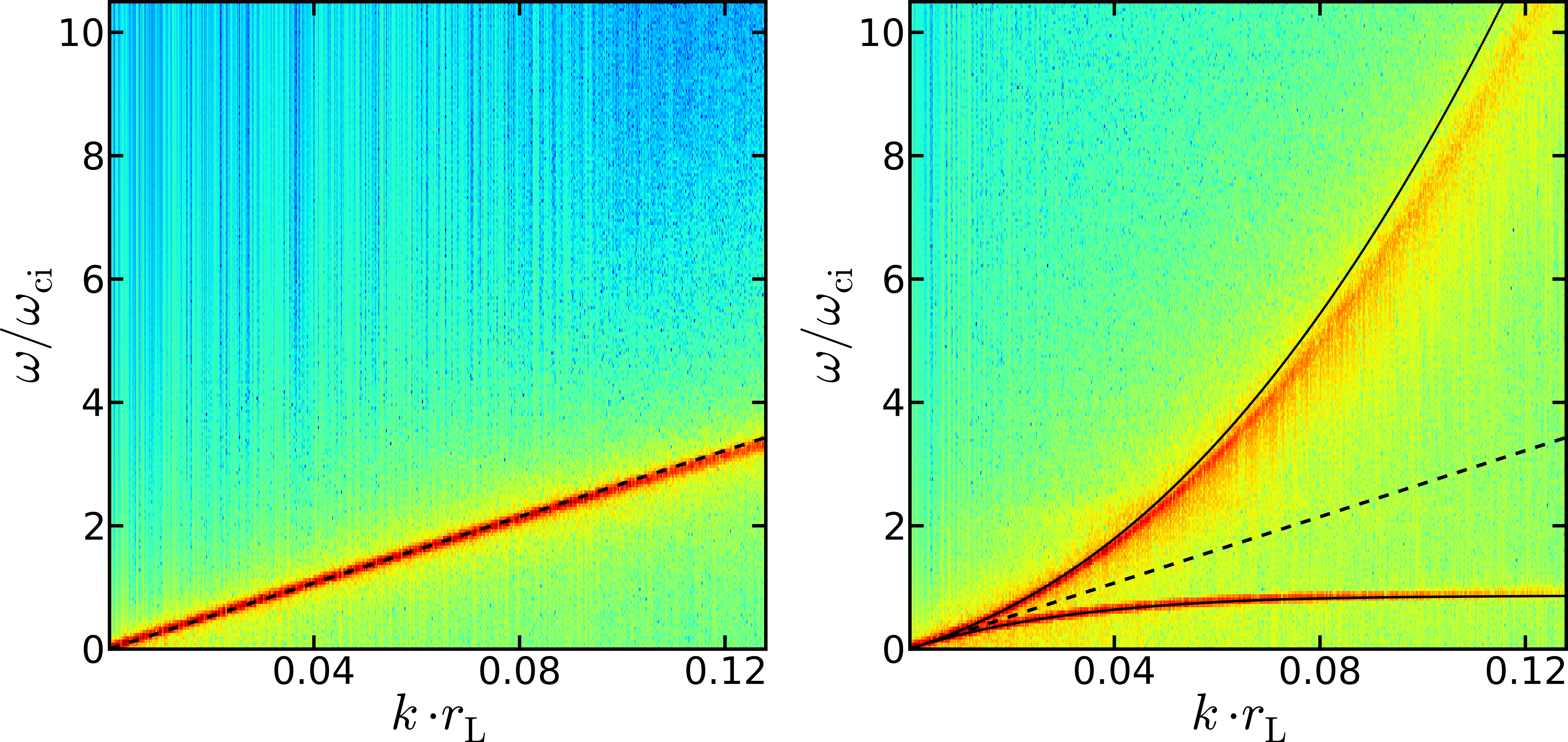}
   \caption{(Color online) Dispersion plot for parallel propagation
   ($\theta=0.001$) with the ideal Ohm's law (left) and Ohm's law including the
   Hall term (right). Further parameters are given in Table
   \ref{tab:Plasma_parameters}. Wavelengths are scaled to the ion gyroradius
   $r_\mathrm{L}$, frequencies to the ion gyrofrequency $\omega_\mathrm{ci}$.
   The dashed line represents the ideal MHD Alfv\'en wave speed
   $v_\mathrm{A}\cos\theta$. The curves represent the WHAMP solution. Without
   the Hall term Vlasiator only shows the non-dispersive ideal MHD Alfv\'en mode.
   With the Hall term in Ohm's law the L-mode resonates at $\omega_\mathrm{ci}$,
   while the higher-frequency branch is the R-mode.}
\label{fig:Par}
\end{figure}

In ideal MHD, Ohm's law takes the form $\mathbf{E}=-\mathbf{V}\times\mathbf{B},$
where $\mathbf{V}$ is the bulk velocity and the Hall term
$\mathbf{j}\times\mathbf{B}/\rho_{q}$ has been omitted with respect to
Equation \eqref{eq:ohms-law}. The absence of spatial scales in this form of
Ohm's law in the hybrid-Vlasov model prevents the kinetic coupling of the ions
to any wave mode. There are thus no resonances and the wave modes present are
the non-dispersive ideal MHD wave modes. In the parallel case this means that the only
wave mode one can observe in the dispersion plot is the ideal MHD Alfv\'en wave
(Eq.~\ref{eq:Disp_MHD_Alfven}), as illustrated in Figure \ref{fig:Par} on the
left-hand side ($\theta=0.001$). All other cases hereafter were obtained using
the Hall term in Ohm's law.

Using $\mathbf{j}=\nabla\times\mathbf{B} / \mu_{0}$ (Eq.\
\ref{eq:maxwell-ampere}) the Hall term can be expressed as
$\left(\nabla\times\mathbf{B}\right)\times\mathbf{B}/\left(\mu_0\rho_{q}
\right)$. It becomes apparent in this form that through the current density,
spatial derivatives of the magnetic field and thus spatial scales are
introduced. One consequence is that wave dispersion can occur in the
hybrid-Vlasov model and one departs from the ideal MHD description.

The dispersion plot for the same parameters but with the Hall term included is
presented on the right-hand side of Figure \ref{fig:Par}. As expected the L-mode
has a resonance at the fundamental ion gyrofrequency. The R-mode on the other
hand is not expected to resonate at $\omega_{\mathrm{ce}}$ (beyond the upper
edge of the figure) because of the lack of electron physics in our hybrid-Vlasov
model. At low frequencies and low wave numbers the two modes converge towards
the ideal MHD Alfv\'en wave. Without the Hall term in Ohm's law, the
simulation result matches the ideal MHD theory. With the Hall term, the simulation
result matches the dispersion obtained by WHAMP. The reason for the slight
discrepancy in the R-mode at higher frequencies is analyzed in Paragraph
\ref{sec:Surf_oblique} below.

\subsection{The ion Bernstein modes in perpendicular propagation}
\begin{figure}
   \centering
   \includegraphics[width=0.5\textwidth]{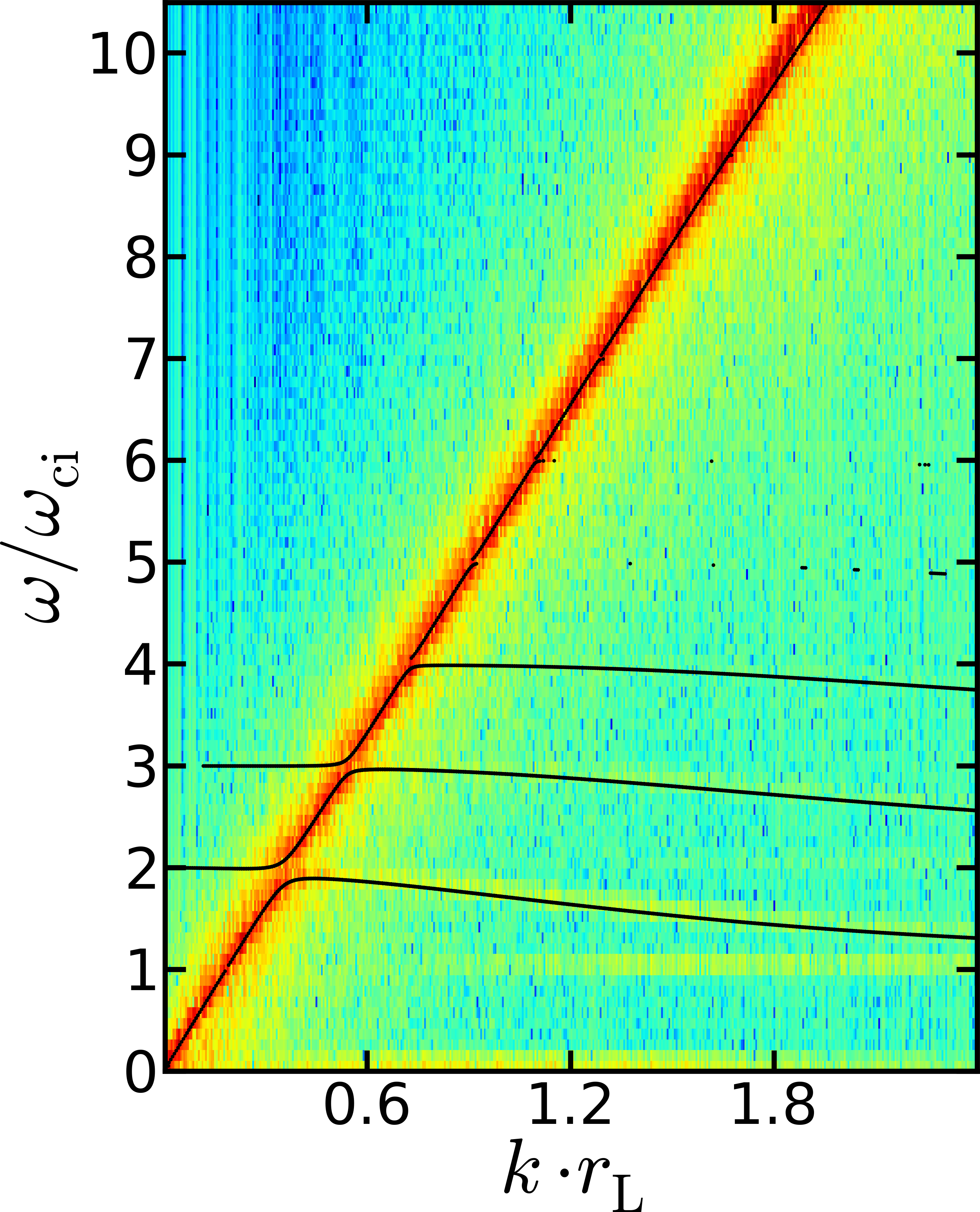}
   \caption{(Color online) Dispersion plot in the case of perpendicular
   propagation ($\theta=1.57$) including the Hall term in Ohm's law. Further
   parameters are given in Table \ref{tab:Plasma_parameters}. The curves
   represent the WHAMP solution. The magnetosonic mode is non-dispersive in
   Vlasiator whereas it bends towards the lower-hybrid plateau in WHAMP (top of
   the plot and beyond), the first few ion Bernstein modes are visible but
   relatively weak.}
\label{fig:Perp}
\end{figure}

In the perpendicular propagation case ($\theta=1.57$) shown in Figure
\ref{fig:Perp} the dispersion plot exhibits the ion Bernstein modes, which are
relatively weak but correspond well to the WHAMP solution. The non-propagating
mode at $\omega_\mathrm{ci}$ and its harmonic at $2\omega_\mathrm{ci}$ are due
to the initial random perturbations fluctuating at the ion gyrofrequency
throughout the simulation domain at all spatial scales. The strongest mode in
the plot is the magnetosonic mode, which is non-dispersive in Vlasiator. It
shows dispersion in the WHAMP solution and bends towards the lower-hybrid
plateau (top of the plot and beyond), a feature not reproduced by Vlasiator
again due to lacking electron physics in the hybrid-Vlasov model.

\subsection{Dispersion surfaces and oblique propagation}
\label{sec:Surf_oblique}
\begin{figure}
   \centering
   \includegraphics[width=\textwidth]{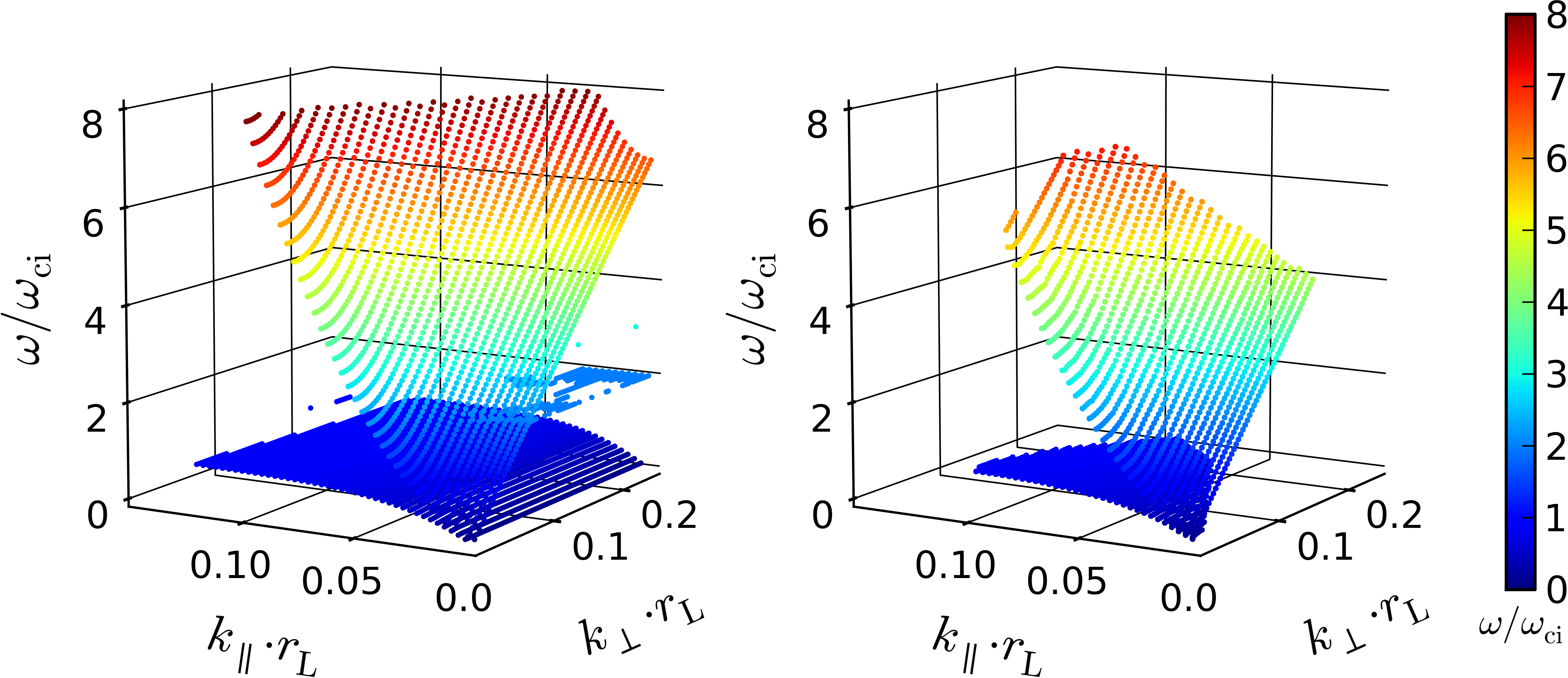}
   \caption{(Color online) Dispersion surfaces for the parameters given in Table
   \ref{tab:Plasma_parameters}. Left: WHAMP results. Right: Vlasiator results
   interpolated and re-sampled from simulations at propagation angles
   $\theta=0.001,0.1,0.3,0.5,0.7,1.0,1.2,1.3,1.4,1.57$. Lower surface: L-mode.
   Higher surface: R-mode. The non-dispersive surface at
   $\omega/\omega_\mathrm{ci}=2$ in the WHAMP solution is the first ion
   Bernstein mode. The WHAMP solution is plotted regardless of the damping of
   the modes.}
\label{fig:Surfaces}
\end{figure}

Following the tradition established by \citet{Andre1985} to study wave
dispersion using dispersion surfaces in the $(k_\perp,k_\parallel,\omega)$
space, we produced the dispersion surfaces for Vlasiator as well. The comparison
of the Vlasiator dispersion surfaces with the WHAMP solution is shown in Figure
\ref{fig:Surfaces}. The WHAMP solution is plotted regardless of the damping or
the relative amplitude of the modes. This explains the smaller extent of the
L-mode surface (lower surface) as well as the absence of the ion Bernstein modes
in this regime in the Vlasiator results.

\begin{figure}
   \centering
   \includegraphics[width=0.5\textwidth]{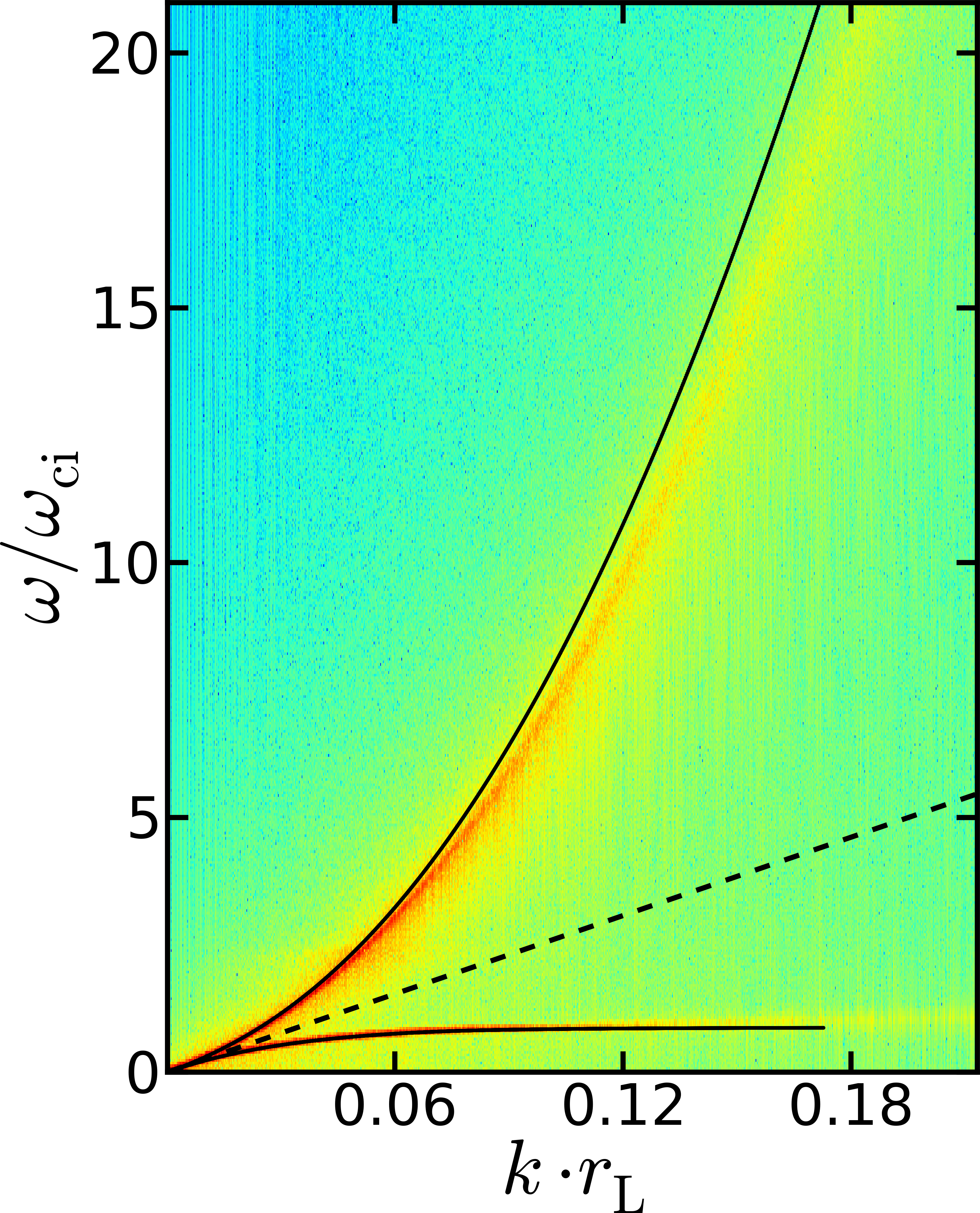}
   \caption{(Color online) Dispersion plot in the case of oblique propagation
   ($\theta=0.3$) including the Hall term in Ohm's law. Further parameters are
   given in Table \ref{tab:Plasma_parameters}. The black curves represent the
   WHAMP solution. The dashed line represents the ideal MHD Alfv\'en wave speed
   $v_\mathrm{A}\cos\theta$.}
\label{fig:Oblique}
\end{figure}

In order to convey a better impression of the relation between the Vlasiator and
WHAMP solutions the dispersion for oblique propagation at $\theta=0.3$ is
presented in Figure \ref{fig:Oblique}. A first observation is that Vlasiator and
WHAMP consistently show a strong damping of the L-mode beyond $k\cdot
r_\mathrm{L}\approx0.18$; the signal weakens in Vlasiator and WHAMP does not
pick the mode any more. Another observation is that while at low frequencies and
long wavelengths both solutions overlap completely, the R-mode does not match
exactly at higher $\omega$, in a similar way as was observed for the parallel
propagation case in Paragraph \ref{sec:Parallel}. Increasing the spatial
resolution of the Vlasiator simulation improves the situation, in that the
Vlasiator R-mode is closer to the WHAMP solution for a longer range. This
indicates that the spatial accuracy of the code, and especially the accuracy of
the Hall term, which is currently of lower order than the Vlasov and field
solvers, is responsible for this discrepancy.

\section{Discussion}
\label{sec:Discussion}
In this study we show that the wave dispersion results from Vlasiator are
closely matched by the WHAMP solution in all propagation directions. The good
correspondence between these two essentially different approaches to solving
Vlasov's equation, namely the linearized kinetic theory and the hybrid-Vlasov
simulation, is an indicator of the good quality of the Vlasiator results. They
clearly show that the Hall term in Ohm's law is critical to go beyond ideal MHD
in terms of wave modes described by the hybrid-Vlasov model. Introducing spatial
scales through the derivatives of the magnetic field, the Hall term in Ohm's law
makes wave dispersion possible. In particular the L- and R-modes split and the
L-mode, coupling to the ion gyromotion, resonates at the ion gyrofrequency. In
addition to this, the ion Bernstein modes are a feature which arises purely from
the kinetic description of magnetized plasma. Two major consequences and
expected shortcomings of the limited electron physics of the hybrid-Vlasov model
are that the R-mode should not resonate at any frequency, and that the
magnetosonic mode does not couple to electrons to form the lower-hybrid plateau
at the ion-electron lower-hybrid frequency $\omega_\mathrm{LH}^{2} =
\left(\omega_\mathrm{pi}^2 + \omega_\mathrm{ci}^2\right) / \left(1 +
\omega_\mathrm{pe}^{2}/\omega_{ce}^2\right)$ in quasi-perpendicular propagation.

This work is a major step in the verification of Vlasiator and it provides
vital insight into the model's wave modes. The aim of the development of
Vlasiator is to provide the first self-consistent hybrid-Vlasov
model able to simulate the entire magnetosphere of the Earth including
ion-kinetic effects. Plasma wave modes are a key feature which should be
described accurately by the model and properly understood in order to interpret
global magnetospheric simulations. It is important to note that the splitting of
wave modes introduced by the Hall term in Ohm's law also occurs on temporal and
spatial scales comparable to and longer than the ion scales. These long scales
are of primary interest in global magnetospheric simulations using a hybrid code
in which the emphasis is placed on ion kinetics. Therefore differences are
expected to appear between Hall-less simulations of the magnetosphere and their
counterparts using the Hall term in Ohm's law, even if the ion gyroradius were
not resolved. Adding the Hall term to Ohm's law in Vlasiator is expected to make
the described physics richer but suitable numerical accuracy and stability in
global magnetospheric simulations will only be achieved with a Hall term of
appropriate spatial accuracy. Realistic results have already been obtained
nevertheless in successful Hall-less magnetospheric simulations, comparing
favorably to observed phenomena\cite{Kempf2013FinCOSPARPresentation}.

\section{Conclusions}
\label{sec:Conclusions}
By studying the dispersion of plasma waves in the hybrid-Vlasov model of the new
Vlasiator code and comparing it to the linearized solution computed with WHAMP,
we provide a key verification benchmark for Vlasiator. The match between the two
fundamentally different approaches is excellent, even with a Hall term in Ohm's
law of first-order spatial accuracy. This study underlines the importance of
using a sufficiently rich Ohm's law in the hybrid-Vlasov model to obtain a
kinetic description of waves departing from ideal MHD, even on temporal and
spatial scales typically associated with ideal MHD. The future addition of
terms to Ohm's law such as the electron pressure gradient term is expected to
play a significant role as well in introducing richer electron physics, the
lower-hybrid plateau for example. This will improve the quality of the model and
the variety of physical phenomena included in it, even more so in the
perspective of implementing adaptive mesh refinement in Vlasiator to resolve
ion-kinetic scales in selected regions.

\section{Acknowledgements}
We acknowledge that the results of this research have been achieved using the
PRACE (Partnership for Advanced Computing in Europe) Research Infrastructure
resource Hermit based in Germany at the High Performance Computing Center
Stuttgart (HLRS).

The Quantifying Energy circulation in Space plasmas (QuESpace) project, in which
Vlasiator is developed, has received funding from the European Research Council
under the European Community's seventh framework programme (FP-7/2007-2013/ERC)
agreement no.\ 200141-QuESpace. The work of YK, DP, SA and MP has been supported
by the Academy of Finland.

\end{document}